\documentclass[prl,twocolumn,showpacs]{revtex4}

\usepackage{graphicx}
\usepackage{amsmath}

\begin{document}

\title{Bose-Fermi Mixtures in a Three-dimensional Optical Lattice}

\author{Kenneth G{\"u}nter, Thilo St{\"o}ferle, Henning Moritz, Michael K{\"o}hl$^*$, Tilman Esslinger }

\affiliation{Institute of Quantum Electronics, ETH Z\"{u}rich,
CH--8093 Z\"{u}rich, Switzerland}

\date{\today}

\begin{abstract}
We have studied mixtures of fermionic $^{40}$K and bosonic
$^{87}$Rb quantum gases in a three-dimensional optical lattice. We
observe that an increasing admixture of the fermionic species
diminishes the phase coherence of the bosonic atoms as measured by
studying both the visibility of the matter wave interference
pattern and the coherence length of the bosons. Moreover, we find
that the attractive interactions between bosons and fermions lead
to an increase of the boson density in the lattice which we
measure by studying three-body recombination in the lattice. In
our data we do not observe three-body loss of the fermionic atoms.
An analysis of the thermodynamics of a noninteracting Bose-Fermi
mixture in the lattice suggests a mechanism for sympathetic
cooling of the fermions in the lattice.
\end{abstract}


\maketitle

Quantum liquids and quantum gases are remarkable objects which reveal
macroscopic quantum phenomena, such as superfluidity and Bose-Einstein
condensation. These fundamental concepts have profoundly influenced our
understanding of quantum many-body physics. The distinct behaviour observed
for purely bosonic or purely fermionic systems sheds light on the role
played by quantum statistics. New insights can be attained by mixing bosonic
and fermionic species. One of the most prominent examples is a mixture of
bosonic $^{4}$He and fermionic $^{3}$He. There it has been observed that
with increasing admixture of $^{3}$He the critical temperature of the
transition between the superfluid and the normal fluid phase is lowered and
below the tricritical point phase separation is encountered \cite{Graf1967}.

In trapped atomic gases mixing of bosonic and fermionic species
has led to the observation of interaction induced losses or
collapse phenomena \cite{Modugno2002,Ospelkaus2006} and
collisionally induced transport in one-dimensional lattices
\cite{Ott2004}. In this work we report on the creation of a novel
quantum system consisting of a mixture of bosonic and fermionic
quantum gases trapped in the periodic potential of a
three-dimensional optical lattice. The optical lattice allows us
to change the character of the system by tuning the depth of the
periodic potential. This leads to a change of the effective mass
and varies the role played by atom-atom interactions. The
interaction between bosonic and fermionic atoms interconnects two
systems of fundamentally different quantum statistics and a wealth
of physics becomes accessible which is beyond that of the purely
bosonic \cite{Jaksch1998,Greiner2002a} or purely fermionic
\cite{Hofstetter2002,Koehl2005} case. A variety of theoretical work
has been devoted to Bose-Fermi mixtures in optical lattices and
new quantum phases have been predicted at zero temperature
\cite{Kuklov2002,Lewenstein2004,Cramer2004,Pollet2005}. Moreover,
the coupling between a fermion and a phonon excitation in the Bose
condensate mimics the physics of polarons \cite{Mathey2004}. At
finite temperature phase transitions to a supersolid state and
phase separation are expected \cite{Buchler2003}.

\begin{figure}[htbp]
  \includegraphics[width=.8\columnwidth,clip=true]{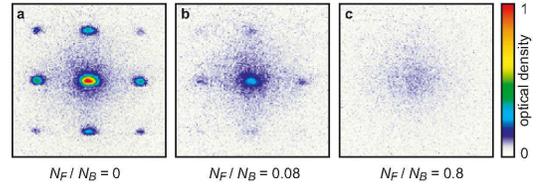}
  \caption{Interference pattern of bosonic atoms released from
           a three-dimensional optical lattice for varying admixture of
           $N_F$ fermionic atoms at a value $U_{BB}/z J_B=5$. The
           bosonic atom numbers are $N_B=1.2\times 10^5$ (a and b) and
           $N_B=8\times 10^4$ (c) and the image size is
           660\,$\mu$m$\times$660\,$\mu$m. The coordination number of our lattice with simple cubic
geometry is $z=6$.}
  \label{fig0}
\end{figure}

In our experiment, we prepare fermionic $^{40}$K atoms together
with a cloud of Bose-Einstein condensed $^{87}$Rb atoms. The
qualitative behaviour when changing the mixing ratio between
bosons and fermions is depicted in figure \ref{fig0}. The momentum
distribution of the pure bosonic sample shows a high contrast
interference pattern reflecting the long-range phase coherence of
the system. Adding fermionic particles results in the loss of
phase coherence of the Bose gas, i.e. a diminishing visibility of
the interference pattern and a reduction of the coherence
length.

Our experimental setup used to produce a degenerate mixture of a
Bose and a Fermi gas has been described in detail in previous work
\cite{Koehl2005}. In brief, fermionic $^{40}$K atoms are
sympathetically cooled by thermal contact with bosonic $^{87}$Rb
atoms, the latter being subjected to forced microwave evaporation.
The potassium atoms are in the hyperfine ground state $|F=9/2,
m_F=9/2\rangle$ and the rubidium atoms in the hyperfine ground
state $|F=2, m_F=2\rangle$. After reaching quantum degeneracy for
both species we transfer both clouds into a crossed beam optical
dipole trap operating at a wavelength of 826\,nm. The laser beams
for the optical dipole trap are aligned in horizontal plane and
their elliptical waists have 1/e$^2$ radii of approximately
50\,$\mu$m and 150\,$\mu$m in the vertical (z) and horizontal
(x,y) directions, respectively. In the optical trap we perform
evaporative cooling by lowering the power in each of the laser
beams to $\simeq 35$\,mW. After recompression, the optical dipole
trap has the final trapping frequencies $(\omega_x,\omega_y,
\omega_z)=2\pi\times (30,35,118)$\,Hz for the rubidium atoms.  We
estimate the condensate fraction to be $90\%$ and use this value
to obtain the temperature of both clouds. The Fermi temperature
$T_F$ in the optical dipole trap is set by the number of potassium
atoms and the trapping frequencies, and we obtain $T/T_F \simeq
0.3$, which is in agreement with a direct temperature measurement
of the fermionic cloud.

The three-dimensional optical lattice is generated by three
mutually orthogonal laser standing waves at a wavelength of
$\lambda=1064$\,nm and a mutual frequency difference of several
10\,MHz. Each of the standing wave fields is focused onto the
position of the quantum degenerate gases and the 1/e$^2$ radii of
the circular beams along the (x,y,z)-directions are
(160,180,160)\,$\mu$m. To load the atoms into the optical lattice
we increase the intensity of the lattice laser beams using a
smooth spline ramp with a duration of 100\,ms. This ensures
adiabatic loading of the optical lattice with populations of
bosons and fermions in the lowest Bloch band only. We have checked
the reversibility of the loading process into the optical lattice
by reversing the loading ramp and subsequently let the particles
equilibrate during 100\,ms in the optical dipole trap without
evaporation. We measure that for both the pure Bose gas and the
Bose-Fermi mixture the condensate fraction decreases by
$\simeq1.4\%$ per $E_R$ of lattice depth, where $E_R=h^2/2
m_{Rb}\lambda^2$ denotes the recoil energy and $m_{Rb}$ the mass
of the rubidium atoms.

The physics of the Bose-Fermi mixture in an optical lattice can be
described by the Bose-Fermi Hubbard model (e.g.
\cite{Lewenstein2004}). The parameters of the model are the
tunnelling matrix elements $J_{B,F}$ for bosons and fermions,
respectively, and the on-site interaction strength $U_{BB}$
between two bosons and $U_{BF}$ between bosons and fermions. Using
the most recent experimental value of the K-Rb s-wave scattering
length \cite{Ferlaino2005} we obtain $U_{BF}/U_{BB} \approx -2$.

\begin{figure}[htbp]
  \includegraphics[width=.75\columnwidth,clip=true]{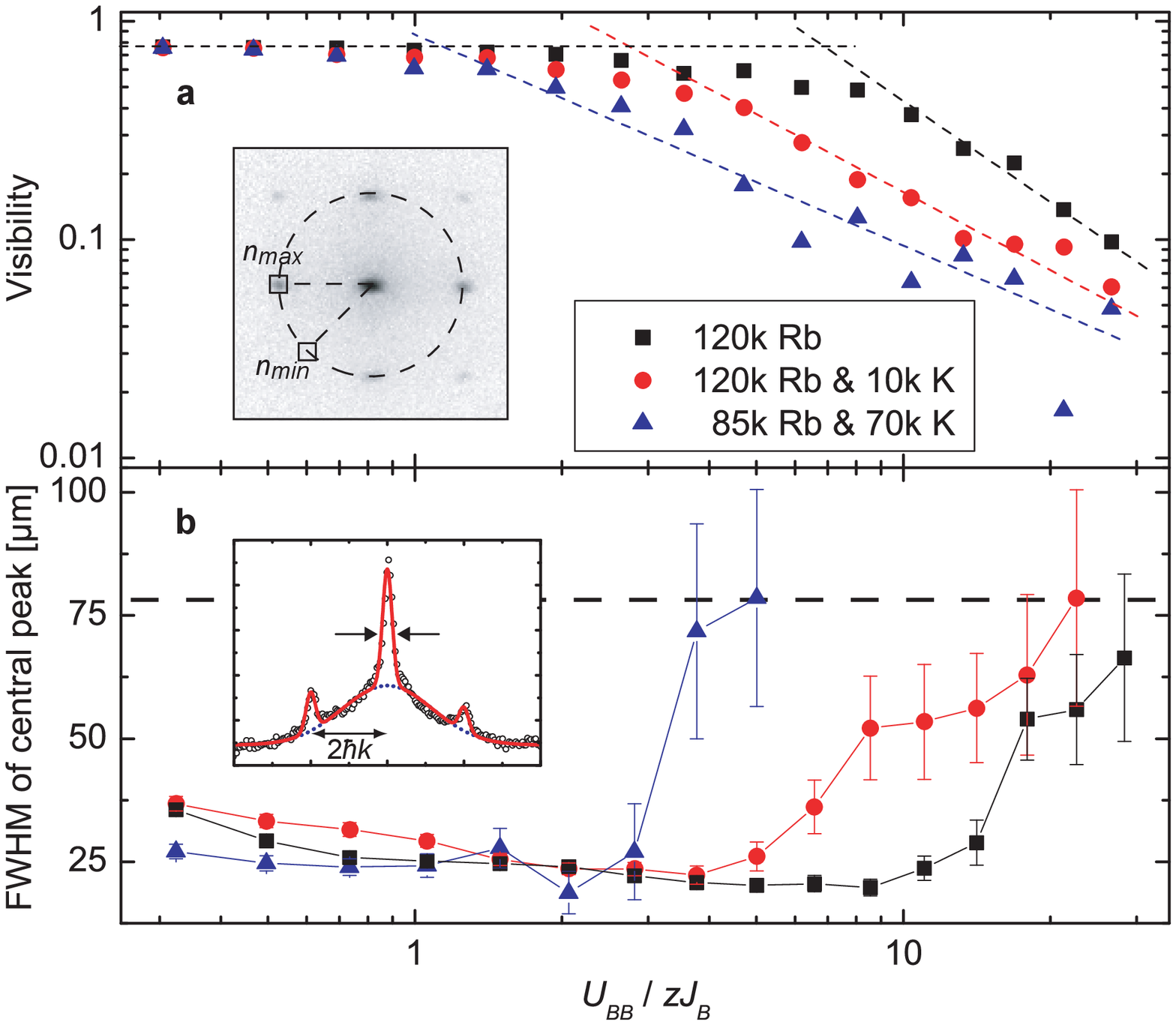}
  \caption{a) Visibility of the Bose-Fermi mixture in the optical lattice for various mixing ratios between bosons and fermions.
  The intersection between the dashed lines defines the characteristic value $(U_{BB}/z J_B)_c$.
  The inset shows the principle of the measurement (see text).
  b) Measurement of the width of the central momentum peak which reflects the inverse of the coherence length of the gas.
  The inset shows how the peak width is extracted from the column sum of the optical density.
  The dashed line indicates the upper constraint of the width imposed by the fitting routine and the error bars reflect the fit uncertainty.}
  \label{fig1}
\end{figure}

We have studied the phase coherence of the bosonic atoms in the
optical lattice for various admixtures of fermionic particles. We
switch off the optical lattice quickly and allow for 25 ms of
ballistic expansion before taking an absorption image of the
atomic cloud. From the absorption image we measure the visibility
of the interference pattern. We determine the maximum $n_{max}$
and the minimum $n_{min}$ of the density of the atoms at a
momentum $|\vec{q}|=2 \hbar k$ with $k=2 \pi/\lambda$ (see inset
in figure \ref{fig1}a) \cite{Gerbier2005}. From this we calculate
the visibility $\mathcal{V}=(n_{max}-n_{min})/(n_{max}+n_{min})$.

For the purely bosonic case we obtain results similar to previous
measurements \cite{Stoeferle2004,Gerbier2005,Xu2006}. In our data (see
figure \ref{fig1}a), the visibility $\mathcal{V}$ starts to drop off at a
characteristic value $(U_{BB}/z J_B)_{c}\approx 6.5$. For larger values of
$U_{BB}/z J_B$ the decrease in visibility is approximated by $\mathcal{V}
\propto (U_{BB}/z J_B)\,^\nu$ with $\nu=-1.41(9)$, which is consistent with
our earlier measurement in a different lattice setup giving $\nu=-1.36(5)$
\cite{Stoeferle2004} but different from the exponent $\nu=-0.98(7)$ obtained
in \cite{Gerbier2005}. For a mixture of bosonic and fermionic atoms the
results change, see also \cite{Bongs2006}. Whereas in the superfluid regime
for very low values of $U_{BB}/zJ_B$ the visibility is similar to the pure
bosonic case, the presence of the fermions decreases the characteristic
value $(U_{BB}/z J_B)_{c}$ beyond which the visibility drops off
significantly. Nevertheless, the visibility still shows a power-law
dependence on $U_{BB}/zJ_B$ with an exponent in the range of $-1<\nu<-1.5$.

\begin{figure}[htbp]
  \includegraphics[width=.65\columnwidth,clip=true]{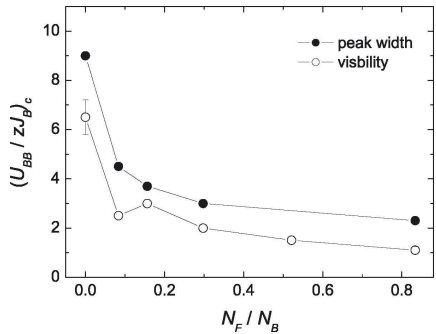}
  \caption{Decrease of the coherence of the Bose gas in the lattice vs. the
  admixture of fermions.}
  \label{fig3}
\end{figure}

To quantify the shift of the visibility data towards smaller
values of $U_{BB}/z J_B$ we have fitted the power-law decay for
large values of $U_{BB}/z J_B$ and extrapolated the slope to the
visibility for the superfluid situation (dashed lines in figure
\ref{fig1}a). The intersection defines the characteristic value
$(U_{BB}/z J_B)_c$ which depends on the mixing ratio between
fermions and bosons $N_F/N_B$ as shown in figure \ref{fig3}. From
this graph it is evident that even very small admixtures of
fermionic atoms change the coherence properties of the bosonic
cloud significantly.

The phase coherence of the bosons in the lattice is not only
described by the visibility of the interference pattern but also
by the coherence length of the sample
\cite{Stoeferle2004,Kollath2004}. In a superfluid state the
coherence length is comparable to the size of the system. The
coherence length is related to the inverse of the width of the
zero-momentum peak plus a small contribution from the repulsive
interaction between the bosonic atoms. For the pure bosonic case
we find that this width starts to increase at a value of
$(U_{BB}/z J_B)_c \approx 9$ (see Fig. \ref{fig1}b). For the case
of a Bose-Fermi mixture this value is dramatically altered: for an
increasing admixture of fermionic atoms to the bosonic sample, the
value decreases and it behaves very similar to the corresponding
value $(U_{BB}/z J_B)_{c}$ for the visibility (see figure
\ref{fig3}).

Our interpretation of the simultaneous decrease of both the coherence length
and the visibility with increasing admixture of fermions is that the system
leaves the superfluid phase. While for the pure bosonic case this indicates
a Mott insulator transition \cite{Stoeferle2004,Gerbier2005}, for the
mixture the analysis is more delicate due to the different interactions and
the different quantum statistics of the two species. The full understanding
of the observed effects including strong interactions and finite temperature
is challenging. We will consider two limiting situations, namely a strongly
interacting Bose-Fermi mixture at $T=0$ in which polarons and composite
fermions are formed, and a non-interacting mixture at finite temperature. In
both explanations we encounter a destruction of the superfluid with
increasing fermionic admixture which qualitatively reflects our results.

At zero temperature several quantum phases of the system are predicted
\cite{Buchler2003,Mathey2004,Lewenstein2004,Cramer2004,Pollet2005},
depending on the sign and the strength of the Bose-Fermi interaction. At low
depth of the optical lattice the interaction of the Bose-Einstein condensate
with the Fermi gas leads to the depletion of the condensate
\cite{Powell2005} and to the formation of polarons where a fermion couples
to a phonon excitation of the condensate \cite{Mathey2004}. The coupling
strength of the fermions to the phonon modes depends on $U_{BF}$ and the
ratio $U_{BB}/J_{B}$. If the coupling becomes very strong the system is
unstable to phase separation ($U_{BF}>0$) or to collapse ($U_{BF}<0$). In
the stable regime, the polarons can form a p-wave superfluid or induce a
charge density wave, as has been analyzed in one spatial dimension
\cite{Mathey2004}. The enhanced bosonic density around a fermionic impurity
increases the effective mass of the fermion and might enhance the tendency
of the bosons to localize. For our parameters, the phonon velocity is
comparable to the Fermi velocity, a regime that is usually unaccessible in
solids. On the other hand, the interaction of the Bose gas with the second
species leads to an effective attractive interaction between the bosons
which would favor a Mott insulator transition at a larger depth of the
optical lattice \cite{Mathey2004}. At a larger depth of the optical lattice
other effects come also into play. Composite fermions consisting of one
fermion and $n_B$ bosons form when the binding energy of the composite
fermion exceeds the gain in kinetic energy that the particles would
encounter by delocalizing. An effective Hamiltonian for these (spinless)
composite fermions with renormalized tunneling and nearest neighbor
interaction has been derived and their quantum phases have been investigated
theoretically \cite{Kuklov2002,Lewenstein2004}. In this situation, the
Bose-Einstein condensate can be completely depleted by the interactions
between bosons and fermions.


For the finite temperature model of the noninteracting gas we
consider the entropy of the cloud of bosons and fermions, which is
$S=\alpha N_F T/T_F + \beta N_B (T/T_c)^3$. $T_F=\hbar
\bar{\omega}_F (6N_F)^{1/3}$ denotes the Fermi temperature for
$N_F$ fermions in a trap with frequency $\bar{\omega}_F$ and
$T_c=\hbar \bar{\omega}_B (N_B/\zeta(3))^{1/3}$ the critical
temperature for Bose-Einstein condensation with $\alpha$ and
$\beta$ being numerical constants. When increasing the depth of
the optical lattice adiabatically, the temperatures of the two
species remain equal to each other due to collisions, while $T_c$
and $T_F$ evolve very differently. This is due to the fact that
the tunnelling rates for the fermions are up to an order of
magnitude larger than for the bosons for our lattice parameters.
Since the effective masses $m^*_{B,F} \propto 1/J_{B,F}$ enter
into the degeneracy temperatures, $T_c$ decreases much faster than
$T_F$. At constant entropy, this results in adiabatic heating of
the bosonic cloud and a reduction of the condensate fraction.
Simultaneously the fermionic cloud is cooled adiabatically,
similar to the situation considered without a lattice in
\cite{Presilla2003}. For the noninteracting mixture with our
parameters one expects a reduction of $T/T_F$ by a factor of
approximately 2 at a lattice depth of $20\,E_R$.

In the experiment we have further studied the occupation of the
optical lattice by measuring three-body recombination. Lattice
sites with a higher occupation than two atoms are subject to
inelastic losses where a deeply bound molecule is formed and
ejected from the lattice together with an energetic atom.
Independent of their occupation all lattice sites are furthermore
subject to loss processes such as off-resonant light scattering,
background gas collisions, or photo-association due to the
trapping laser light. The attractive interaction between the
bosons and the fermions changes the occupation of bosons on the
sites of the optical lattice. For the given ratio of the on-site
interaction strength of $U_{BF}/U_{BB} \simeq -2$ it is
energetically favorable to have up to five bosons per site if a
fermion is present.

\begin{figure}[htbp]
  \includegraphics[width=.7\columnwidth,clip=true]{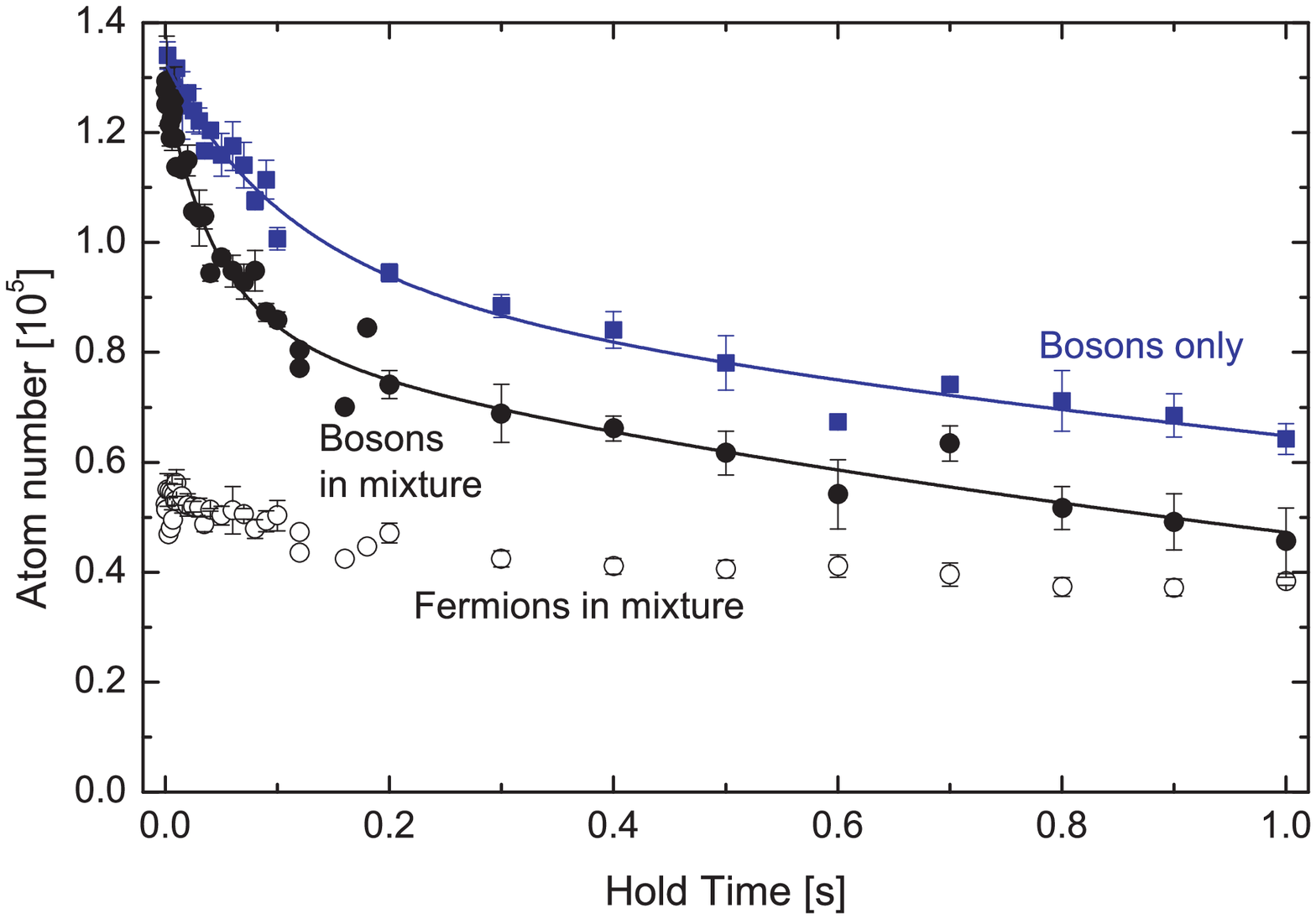}
  \caption{Decay of a pure bosonic gas (squares) and a Bose-Fermi mixture (circles) in the optical lattice.
  The fast initial decay of the bosons is much more pronounced in the mixture,
  reflecting the higher density due to Bose-Fermi attraction.
  For the fermions hardly any loss is observed. The error bars indicate statistical errors from three repetitive measurements.}
  \label{fig4}
\end{figure}

The experimental sequence to study the three-body decay starts
from an initially superfluid Bose gas at a potential depth of
10\,$E_R$.  We use a ramp time of 30\,ms to increase the potential
depth of the lattice from zero to 10\,$E_R$ during which we do not
observe a loss of atoms. Subsequently, we freeze the atom number
distribution by quickly changing the lattice depth to a large
value of 18\,$E_R$ where the tunnelling time of the bosons is
$\tau_B=1/zJ_B=23$\,ms. We monitor the total atom number as a
function of the hold time in the deep optical lattice (see figure
\ref{fig4}) and observe two distinct time scales of the decay of
the atoms. The fast initial time scale is due to three-body losses
from multiply occupied lattice sites. The slower decay is due to
single-particle loss processes.

To extract quantitative information from the loss curves we fit
the data with the following model. We assume that any singly or
doubly occupied site decays with a single particle loss rate
$\Gamma_1$. Multiply occupied sites decay with a rate determined
by the three-body loss constant $K_3^B=1.8\times
10^{-29}$\,cm$^{6}$/s \cite{Soding1999} and the three-body density
$[n({\bf r})]^3$ at the lattice site. Since we start from a
superfluid the number distribution at the lattice sites can be
approximated by a coherent state with a third order correlation
function being equal to unity \cite{Burt1997}. We calculate the
three-body loss rate assuming gaussian ground state wave functions
at each lattice site to be $\Gamma_3=0.24\times n_B^3$\,s$^{-1}$,
where $n_B$ is the number of bosons on the site. By fitting the
data with this model we extract the occupation of the lattice. We
obtain $n_{1,2}=67(3)\%$ of the sites with single or double
occupation, $n_{3}=23(9)\%$ sites with triple occupation and
$n_4=10(8)\%$ of lattice sites with occupation four. A mean field
calculation neglecting tunnelling yields the theoretical values
$\bar{n}_{1,2}=58\%, \bar{n}_3=33\%$, and $\bar{n}_4=17\%$ which
gives reasonable agreement given the simplicity of the model. The
slow decay rate is determined to be 0.35(7)\,s$^{-1}$.

Upon adding fermions to the system we find a much faster initial
decay due to three-body loss for the rubidium atoms. The single
particle loss constant is, however, the same. In contrast, for the
fermionic atoms we do not observe a particle loss of a comparable
order of magnitude. This suggests that the observed loss is only
due to three-body recombination between three rubidium atoms.
Recent results  have suggested that the three-body loss constant
$K_3^{BF}$ for K-Rb-Rb collisions is an order of magnitude larger
than for Rb-Rb-Rb collisions \cite{Ospelkaus2006}. This is not
consistent with our data since we do not observe the corresponding
fast loss of potassium atoms, similar to previous results
\cite{Goldwin2004}.

In conclusion, we have investigated a Bose-Fermi mixture in a
three-dimensional optical lattice. We have observed that the
presence of fermions changes the coherence properties of the Bose
gas and substantially enhances the three-body loss of bosonic
atoms. Bose-Fermi mixtures in an optical lattice promise to be an
incredibly rich quantum system
\cite{Kuklov2002,Lewenstein2004,Cramer2004,Mathey2004,Buchler2003,Pollet2005}.
A number of Feshbach resonances between $^{87}$Rb and $^{40}$K
\cite{Inouye2004,Ferlaino2005} exists which will give access to
various quantum many-body regimes predicted in the literature as
well as to the creation of ultracold heteronuclear molecules.

We would like to thank G. Blatter, W. Hofstetter, C. Kollath, A.
Kuklov, L. Pollet, M. Troyer for insightful discussions, and
OLAQUI, SNF and SEP Information Sciences for funding.

\end{document}